\documentclass[lettersize,journal]{IEEEtran}    
\usepackage{amsmath,amsfonts}
\usepackage{algorithmic}
\usepackage{algorithm}
\usepackage{array}
\usepackage[caption=true,font=normalsize,labelfont=sf,textfont=sf]{subfig}
\usepackage{textcomp}
\usepackage{stfloats}
\usepackage{url}
\usepackage{verbatim}
\usepackage{graphicx}
\usepackage{cite}
\usepackage{flushend}
\usepackage{bm}
\usepackage[czech]{babel}
\usepackage[T1]{fontenc}
\usepackage[utf8]{inputenc}

\usepackage[colorlinks,allcolors=black]{hyperref}
\begin{document}

\title{Magic NeRF Lens: Interactive Fusion of Neural Radiance Fields for Virtual Facility Inspection}

\author{Ke Li, Susanne Schmidt, Tim Rolff, Reinhard Bacher, Wim Leemans, Frank Steinicke 

\thanks{ Ke Li, Reinhard Bacher, and Wim Leemans are with the accelerator division at the Deutsche Elektronen Synchrotron DESY, Notkestraße 85, 22607 Hamburg, Germany. Email: ke.li1@desy.de}

\thanks{Ke Li, Susanne Schmidt, Tim Rolff, and Frank Steinicke are with the Human Computer Interaction group at Hamburg University, Vogt-Kölln-Straße 30, 22527 Hamburg, Germany.}

\thanks{Tim Rolff is with the Computer Vision group at Hamburg University, Vogt-Kölln-Straße 30, 22527 Hamburg, Germany.}

\thanks{Please find in the supplementary material the demo video illustrating our contributions. }
}



\IEEEpubid{}

\maketitle

\begin{abstract}
Large industrial facilities such as particle accelerators and nuclear power plants are critical infrastructures for scientific research and industrial processes. These facilities are complex systems that not only require regular maintenance and upgrades but are often inaccessible to humans due to various safety hazards. Therefore, a virtual reality (VR) system that can quickly replicate real-world remote environments to provide users with a high level of spatial and situational awareness is crucial for facility maintenance planning. However, the exact 3D shapes of these facilities are often too complex to be accurately modeled with geometric primitives through the traditional rasterization pipeline. 

In this work, we develop Magic NeRF Lens, an interactive framework to support facility inspection in immersive VR using neural radiance fields (NeRF) and volumetric rendering. We introduce a novel data fusion approach that combines the complementary strengths of volumetric rendering and geometric rasterization, allowing a NeRF model to be merged with other conventional 3D data, such as a computer-aided design model. We develop two novel 3D magic lens effects to optimize NeRF rendering by exploiting the properties of human vision and context-aware visualization. We demonstrate the high usability of our framework and methods through a technical benchmark, a visual search user study, and expert reviews. In addition, the source code of our VR NeRF framework is made publicly available for future research and development \footnote{\url{https://github.com/uhhhci/immersive-ngp}}. 

\end{abstract}


\section{Introduction}
\noindent In recent decades, an increasing number of new large-scale facilities have been built around the world. These facilities, including particle accelerators, nuclear power plants, and lithography machines, are critical infrastructures for scientific research and industrial processes. As they are often complex systems built after decades of planning and implemented at enormous economic cost, scientists and engineers today are actively seeking methods to effectively maintain existing facilities in order to maximize their operational lifetimes and economically upgrade them to new operational standards. Unfortunately, many of these facilities are hazardous environments that are inaccessible to humans most of the time. For example, at the Large Hadron Collider (LHC) at the European Organization for Nuclear Research (CERN), the particle accelerator could generate high-energy gamma radiation at doses lethal to humans during operation, making on-site operation of the accelerator tunnel limited and expensive \cite{50-CERNBot}. As a result, a virtual inspection system where accurate 3D representations of the complex conditions of the facility can be viewed and updated frequently is critical for maintenance planning to reduce maintenance windows. 

\begin{figure}
    \centering
    \includegraphics[width=1\linewidth]{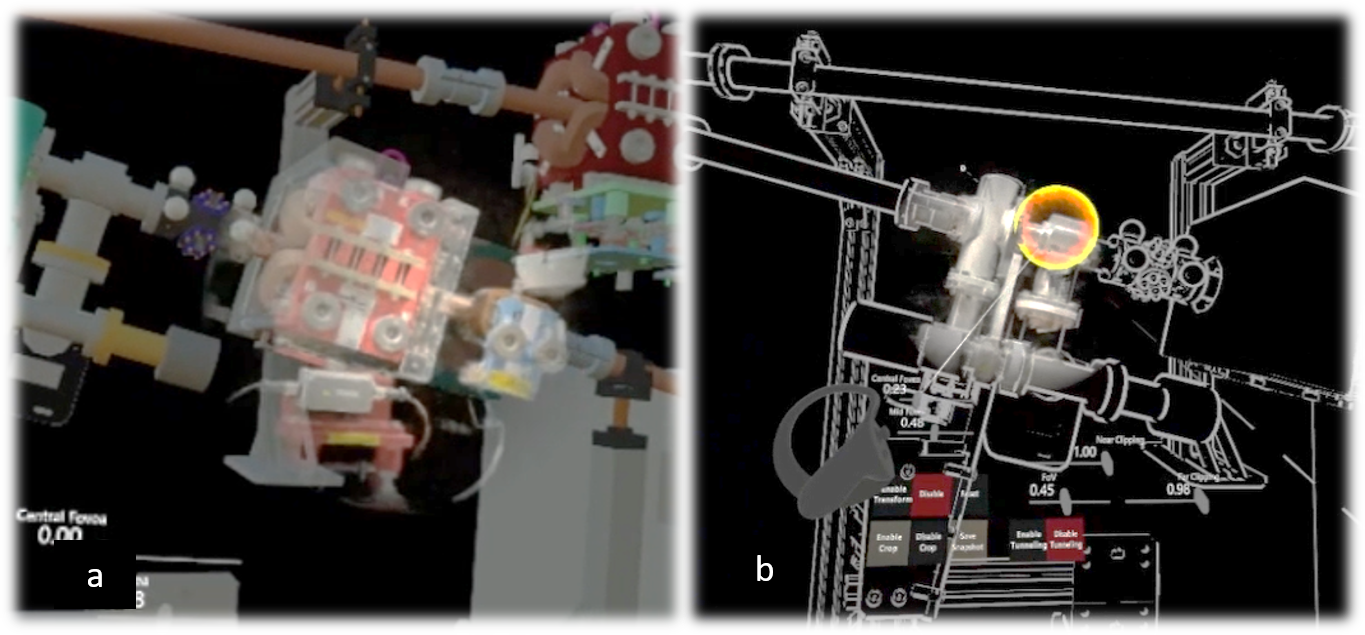}
    \caption{ Screenshots from the Magic NeRF Lens framework: (a) illustrates the data fusion of a high-resolution NeRF rendering and corresponding CAD model through mixed reality tunneling \cite{19-MR-Tunneling}, and (b) illustrates the 3D NeRF drawing interaction using the CAD model as context. }
    \label{fig:teaser}
\end{figure}

In fact, how to accurately replicate such complex real-world environments and how users could virtually interact with their digital replica is at the heart of computer graphics research. On the one hand, the exact 3D shapes of detailed components are too complex to be accurately modeled, updated, and rasterized using traditional geometric representations such as computer-aided design (CAD) models. On the other hand, off-the-shelf RGBD sensors \cite{98-star-rgbd} or photogrammetry methods \cite{99-heritage} have limited capabilities to solve the challenging inverse 3D reconstruction problems, while professional 3D reconstruction pipelines, such as those used in heritage preservation processes, require expensive hardware and involve time-consuming multimodal data fusion and refinement \cite{83-digital-preservation, 99-heritage}. 


In recent years, neural radiance fields (NeRFs) \cite{26-Original-NeRF} have emerged as a novel photorealistic 3D representation that can be generated by a neural network learning only from 2D images and their camera poses. In particular, the development of Instant Neural Graphics Primitives (instant-ngp) \cite{27-Instant-NGP} is generating great excitement in the community, as it drastically accelerates NeRF training and rendering speed through an efficient multi-resolution hashing data structure. However, unlike traditional rasterization pipelines, a NeRF generates 2D projections of 3D space through volumetric ray casting, whose performance scales largely with the rendering resolution. This hinders its application in real-time, stereoscopic, high resolution, high frame rate rendering in immersive virtual reality (VR).  As a result, research on VR-NeRF interactive systems and applications is currently sparse. In addition, the question of how to interact with such neural 3D volumetric representations and how to integrate them into the conventional rendering pipeline is still under active research \cite{94-Neural-rendering-general}. 

In this paper, we present Magic NeRF Lens, a virtual facility inspection framework that could exploit the complementary features of neural graphics primitives such as NeRF and conventional geometric primitives such as a CAD model through data fusion. In particular, our framework enables facility inspection in immersive VR, which is the emerging technology that is widely believed to become how users will ultimately interact with 3D spatial information in the future, and has already demonstrated advantages in terms of improving users' spatial presence \cite{071-Spatial-perception-CAD} and spatial skills in such facility inspection and management tasks \cite{11-AR-2D-3D-Comparison,  07-compare-CAD-VR-2D}. 

A common requirement for virtual inspection in immersive VR is to render 3D scenes at one-to-one real-world scale to provide users with correct spatial and situational awareness of the remote environments \cite{89-multi-modal-CERN,51-multi-model-CERN, 82-VR-factory-point-cloud, 36-HRI-Li}. Since rendering these NeRF models in today's wide field-of-view (FoV) VR head-mounted displays (HMDs) requires enormous computational resources, and real-time performance is only feasible by significantly degrading rendering resolution, we introduce two novel contextual volume data visualization and interaction techniques to balance this trade-off between rendering performance and visual quality. As Figure \ref{fig:teaser} (a) illustrates, the first data fusion design employs a ``mixed reality (MR) tunneling'' effect \cite{19-MR-Tunneling}. Here, the high-resolution NeRF rendering is displayed in the center of the user's field of view and fuses with the CAD model displayed at the periphery. As shown in figure \ref{fig:teaser} (b), the second design allows the user to define the target rendering areas by directly manipulating a binary volume bitfield via 3D drawing, using the CAD model as a contextual guide. 

We validate the effectiveness of our methods and framework through technical benchmarking and a formal study in which users performed visual search tasks for facility inspection. We conducted confirmatory expert reviews with five particle accelerator control and management specialists at the German Electron Synchrotron (DESY), where some of the world's most advanced particle accelerators are built. Finally, we are releasing the source code for our VR-NeRF system, which enables features such as object manipulation, volumetric scene editing, and depth occlusion effects in the popular Unity game engine \cite{28-Immersive-NGP, 93-interactive-nerf}. As the first open source VR NeRF system available, we believe it will have a high impact on computer graphics and human-computer interaction (HCI) research. 

In summary, the contribution of this work includes:

\begin{enumerate}
    \item A VR-NeRF framework for virtual industrial facility inspection. 
    \item Two novel contextual data fusion designs to support interaction and rendering of industrial facility NeRF models at real-world size in immersive VR.
    \item Comprehensive evaluation of the framework and data fusion designs through technical benchmarking, a user study, and expert reviews. 
    \item An open-source VR-NeRF system with features such as NeRF object manipulation, volumetric editing, and depth occlusion, all integrated into the Unity game engine.
\end{enumerate}

\section{Related Work}


\subsection{Large-scale Industrial Facility Maintenance}

\noindent Large industrial facilities such as the LHC at CERN \cite{85-LHC} are well known for their scientific missions. However, they also present some of the most demanding HCI and computer graphics challenges in facility control and visualization. One of the most significant challenges is planning and completing a large amount of maintenance and inspection work during very limited on-site access windows.
For example, the 2.3-kilometer-long particle accelerator at the European X-Ray Free Electron Laser (EuXFEL) is a complex system with 10 million control system parameters and tens of thousands of components that require frequent inspection and maintenance. However, the accelerator must operate continuously for more than 5,000 hours per year, during which time on-site human access is not possible and any unexpected interruptions to operation result in high energy and setup costs \cite{96-MARWIN}. To document the facility status for offline inspection and planning, Gong et al. propose a VR system by mixing CAD models with updated LiDAR 3D point cloud to visualize the real factory layout \cite{82-VR-factory-point-cloud}. Bae et.al. propose an augmented reality (AR) system for contextual asset management by mixing 2D images with historical 3D data \cite{08-3Dpointcloud-mobile-phone-2d-overlay}. There is also extensive research into the use of telepresence and teleoperated mobile robots in such environments. However, much research is needed before robots can operate autonomously without human intervention for tasks that require fine motor skills \cite{50-CERNBot, 52-AR-CERN-Bot,89-multi-modal-CERN}.
Regardless of virtual planning or robot teleoperation in VR, a photorealistic 3D model of the real facility conditions can greatly improve the various workflows of managing these facilities.

\subsection{NeRF}

\noindent NeRF is widely regarded as a groundbreaking advance in 3D computer graphics and has received tremendous attention since its initial development \cite{26-Original-NeRF}. By training a small neural network with only 2D images and their camera poses, a NeRF can compress scene details into a small network model instead of storing all the color values of the entire voxel grid on disk. Compared to photogrammetry or conventional RGBD sensors, creating a NeRF does not require bundle adjustment and dense reconstruction, but could produce higher visual quality than photogrammetric point clouds, which tend to be erroneous with limited image feature overlap or non-uniform textures \cite{98-star-rgbd}. 
Compared to active 3D scanners with sub-millimeter accuracy such as those with structured illumination \cite{100-sl-review}, NeRF can work with ``optically uncooperative" surfaces such as metallic or absorbent materials that are common in industrial facilities. While it is also common to combine both image-based and active 3D scanning to create a high-quality 3D digital twin \cite{99-heritage}, it is noteworthy that the post-processing associated with these methods can take weeks to months, during which time the conditions of the facility may have already changed due to maintenance activities. 
As a result, implicit 3D models such as NeRF are a compelling alternative to traditional meshes or point clouds for rapid documentation of facility conditions.  

\subsection{VR NeRF}
The rapid progress in NeRF training \cite{26-Original-NeRF, 27-Instant-NGP, 34-block-nerf}, rendering \cite{27-Instant-NGP, 30-mip-nerf, 35-fov-nerf}, and editing \cite{31-control-nerf, 32-instructnerf2023, 33-NeRFshop23} is also causing great excitement in the VR community. With the availability of real-time NeRF rendering techniques such as instant-ngp \cite{27-Instant-NGP}, the question of how to complement NeRF rendering with practical user interfaces to support different application domains also becomes relevant for HCI research. However, the availability of open source VR NeRF systems remains sparse. 
In fact, most open source systems today only support scene visualization on 2D desktop \cite{38-nerf-studio,35-fov-nerf,33-NeRFshop23 } or were only available as proof-of-concept implementations using high-level programming languages such as Python, which has poor interoperability with popular graphics rendering engines. To fill this gap, we released immersive-ngp \cite{28-Immersive-NGP}, the precursor to this work, presenting the first software for viewing NeRF scenes in a VR HMD integrated with the Unity game engine. Although instant-ngp subsequently introduced the ability to render a NeRF scene on a VR-HMD, the scalability of the instant-ngp VR system is limited, as integrating additional 3D models into the original framework would require users to rebuild the entire interaction and rendering pipeline from scratch \cite{27-Instant-NGP}. This work provides further extensions to immersive-ngp so that users can easily manipulate, edit, and integrate NeRF into their own interactive 3D applications. \\

In addition to the lack of open source VR NeRF systems, another challenge for VR NeRF is the enormous amount of network queries required for stereoscopic, high resolution, high frame rate VR rendering \cite{95-steerNeRF}. One promising direction is to use foveated rendering to reduce render resolution in the peripheral region of the human visual field. However, the existing foveated NeRF method requires training and recombining the rendering results of separate networks. 
Moreover, it did not use the efficient multi-resolution hash coding data structure, which can achieve most of the rendering and performance speedup \cite{35-fov-nerf}. This work provides an alternative method to conventional foveated rendering by displaying a simpler 3D representation at the peripheral region of the human eye rendered through rapid geometric rasterization. Our method offers a more scalable solution for performance optimization and provides an example for the current debate on how NeRF could be integrated into the conventional graphics framework.

\subsection{Magic Lens}

\noindent The magic lens effect was first developed by Bier et al. \cite{20-magic-lens-orginal}, allowing users to change the visual appearance of a user-defined area of the user interface by overlaying a transparent lens over the render target. Interactive magic lens techniques are widely used in modern visualization systems \cite{23-survey-interactive-lense}, where context-aware rendering of large information spaces is needed to save computational costs \cite{24-the-magic-volume-lens,21-3D-Magic-Lens}. In 3D computer graphics, several 3D magic lens effects have been developed to allocate computational resources to more resolution-important features for the visualization of volumetric medical scans \cite{24-the-magic-volume-lens,21-3D-Magic-Lens} or context-aware AR applications \cite{91-AR-magic-lens, 92-handheld-magic-lens-ar}. 
For immersive VR HMD rendering, the MR tunneling effect is also an interactive lensing technique that can exploit the characteristics of human vision to merge data from two sensors with different resolutions, frame rates, and latency \cite{19-MR-Tunneling}. This work extends the conventional 3D magic lens techniques to NeRF rendering by reducing the number of rays and sampling queries to the neural network.

\section{The Magic NeRF Lens Effects} \label{sec:magic-nerf-lens-theory}

\subsection{VR NeRF Rendering Formulation} 

\noindent As suggested by Mildenhall et al. \cite{26-Original-NeRF}, NeRF represents the 3D world with a volumetric scene function $F_{\Theta}$ with learnable parameters $\Theta$ by mapping the 3D position $\bm{p}(x,y,z)$ and the viewing direction $d(\theta, \phi)$ to a color emission vector $\bm{c}(r,g,b)$ and a volume density float $\sigma$: 
\begin{equation}
    F_{\Theta} : (x,y,z, \theta,\phi) \mapsto (r,g,b, \sigma).
\end{equation}
To generate a 2D projection from this 3D scene representation, NeRF rendering relies on conventional volumetric ray-casting rendering. A camera ray $\bm{r}_{(x,y)}$ from $\bm{o}(x,y)$ with viewing direction $d$ is cast into 3D space for each pixel $(x,y)$: $\bm{r}_{(x,y)}(t) = o(x,y) + t \cdot d$, where $t$ is within the interval of the user-defined near boundary $t_{n}$ and far boundary $t_{f}$ of the ray: $t \in [t_{n}, t_{f}]$. 

The color value $\bm{C}(\bm{r})$ per pixel can be calculated by sampling and accumulating the transmittance along the ray $T(t)$, the volume density function $\sigma(t)$, and the color value $\bm{c}(t)$:

\begin{equation}
     C(\bm{r}) = \int_{t_\text{n}}^{t_{\text{f}}} T(t)\cdot \sigma(\bm{r}(t))\cdot c(\bm{r}(t),d)\ dt.
\end{equation}

As a result, the NeRF rendering performance $P_{(h,w)}$ is bounded by the target render resolution with $R_h$ height, $R_w$ width, the number per ray $N_{r(t)}$, and the average speed of each network query (usually measured in FLOPs) $\overline{F}$ \cite{95-steerNeRF}:  

\begin{equation}
    P_{(h,w)} = R_h \times R_w \times N_{\bm{r}(t)} \times \overline{F}
\end{equation}

Although the multi-resolution hash encoding data structure of instant-ngp \cite{27-Instant-NGP} heavily parallelizes the network query process and achieves rendering speedup by significantly reducing $\overline{F}$, another strategy to further optimize rendering performance is to reduce the large number of feed-forward network queries $N_{\bm{r}(t)}$. One way to achieve this is to reduce the overall render resolution $R_{(h,w)}$, which will inevitably lead to a degradation of the rendering visual quality. 

However, for a VR HMD, the per-eye NeRF rendering performance $P_{hmd}$ is determined by the horizontal field of view $FoV_{h}$, the vertical field of view $FoV_{v}$ of the HMD, and the pixel density per degree (PPD), which is the intrinsic parameter of the display hardware: 

\begin{equation} \label{eq:hmd-p}
    P_{hmd} = FoV_{h} \times FoV_{v} \times PPD^{2} \times C \times N_{\bm{r}(t)} \times \overline{F} .
\end{equation}

Note that the constant $C$ is a multiplication factor, given the supersampling requirements in practice for aliasing reduction and anti-aliasing \cite{79-supersampling}.

From equation \ref{eq:hmd-p} we could see that reducing the FoV of NeRF rendering in VR could also result in a significant speedup without sacrificing visual quality. Another way to optimize $P_{(h,w)}$ is to reduce the overall sample size $N_{\bm{r}(t)}$. Abundant previous work has shown that this can be achieved by empty space skipping and early ray termination techniques \cite{97-NeRFAcc}, which were already part of the original instant-ngp implementation \cite{27-Instant-NGP}. In the remainder of this section, we demonstrate the design of two 3D magic lens effects to efficiently utilize the performance optimizers, including FoV restrictor, empty space skipping, and early ray termination, for rendering real-world scale industrial facilities in a VR HMD.

\subsection{Magic NeRF Lens with FoV Restrictor}

\begin{figure}
    \centering
    \includegraphics[width=1\linewidth]{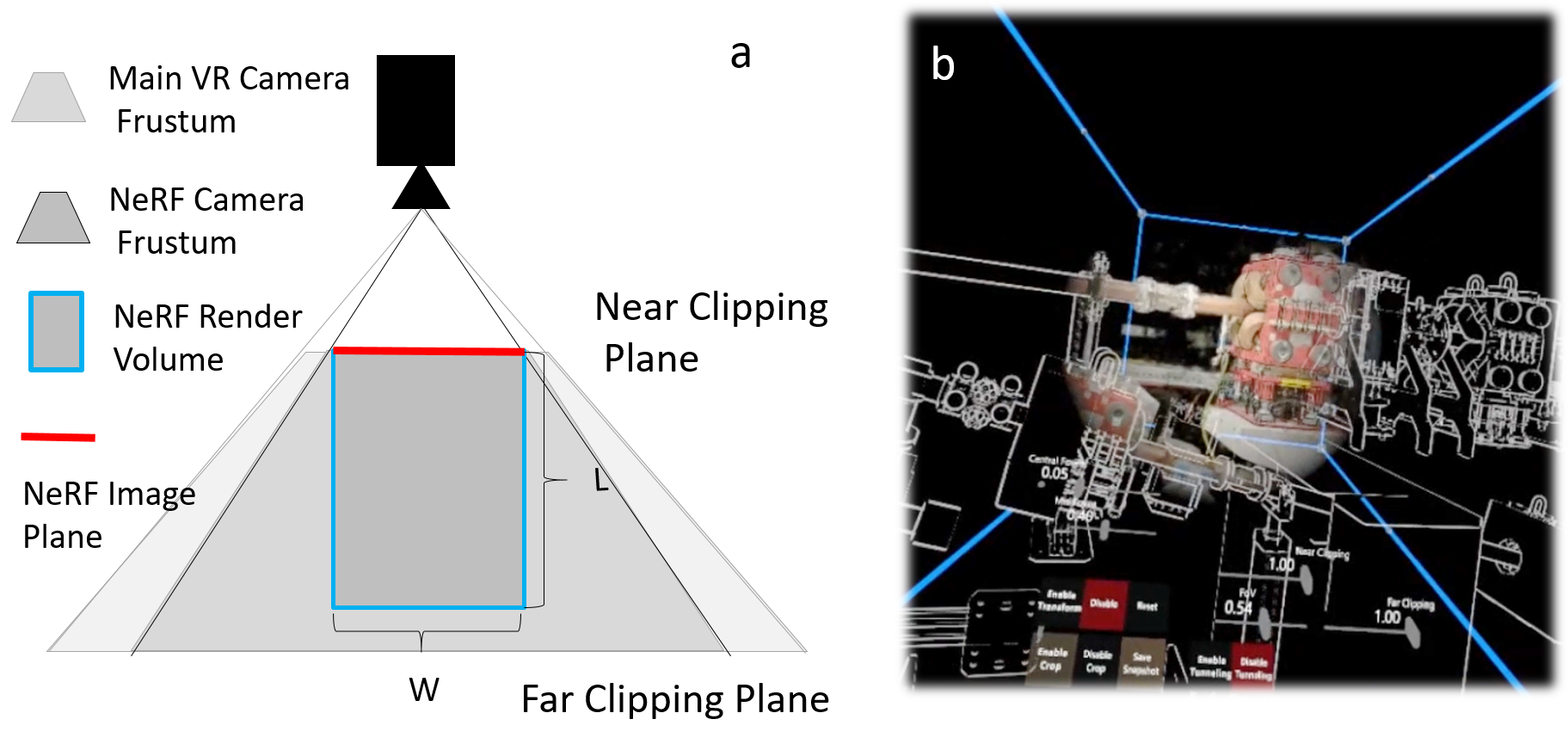}
    \caption{(a): Schematic sketch of the design of our interactive lens effect where the FoV of the NeRF camera is reduced and the actual NeRF render frustum is defined as a box rather than a pyramid to reduce NeRF render load.  (b): Screenshot of the magic lens effect, where the blue box visualizes the NeRF crop box that is dynamically following the user's head movement. }
    \label{fig:tunneling_nerf}
\end{figure}

By applying a FoV restrictor to the NeRF model, it creates a ``VR tunneling'' effect \cite{58-FoV-restrictor}, which is a typical motion sickness reduction technique in VR gaming to restrict the optical flow and sensory conflicts of the peripheral region of the human eye \cite{59-FoVRestrictor2001}. However, a FoV restrictor could significantly reduce the user's sense of presence and immersion. To compensate for the NeRF rendering reduction, we implement the ``MR tunneling" effect \cite{19-MR-Tunneling} by displaying a CAD model or other types of conventional 3D representations of the industrial facilities in the peripheral regions of the user's field of view. Previous research shows that MR tunneling could save computational costs by merging two visual modalities with different resolution, frame rate, and motion-to-photon latency while achieving more desirable overall perceptual effects \cite{19-MR-Tunneling}. 

Figure \ref{fig:tunneling_nerf}(a) shows the schematic relationship between the main VR camera, the NeRF camera, and the active NeRF rendering volume. Figure \ref{fig:tunneling_nerf} (b) shows the volumetric crop box of the NeRF model, which is dynamically aligned with the user's viewing direction and the near clipping plane of the VR camera, so that the volume crop box acts as an interactive lens that automatically selects the target rendering volume according to the size of the image plane $W$ and the distance of the far clipping plane $L$. The parameters $W$ and $L$ are both user-defined values to reduce the total number of sampling points and could be adjusted during application runtime to avoid large frame rate jitter. For the static MR tunneling effect, the high-resolution NeRF rendering is displayed in the central region of the HMD. However, an eye tracker could be integrated to achieve foveated MR tunneling and reduce the need for frequent head movements while inspecting the facilities \cite{19-MR-Tunneling}.   

\subsection{Magic NeRF Lens with Context-aware 3D Drawing} \label{sec:3D_NeRF_Drawing}

\noindent For some facility inspection tasks, users typically do not need the entire NeRF model for their tasks. For example, components such as walls, simple electrical boxes, floors, and ceilings are static elements that normally do not need to be inspected and modified, and could be visualized simply by their CAD models. Based on the original idea of magic lens in user-defined context-aware rendering, we design the magic NeRF lens effect with 3D drawing interaction, where users can dynamically reveal and erase an area of the NeRF model. As shown in Figure \ref{fig:teaser} (b), the wireframe representation of a CAD model outlines the edges and general context of the environment, and users could adjust the radius of a 3D sphere attached to the VR controller and point the 3D sphere at a spatial location where the NeRF render volume should be revealed on demand. 

The transparency representation of NeRF could be considered as a 3D density grid $\sigma(h,w,l)$, where $h$, $w$, $l$ are the data height, width, and length. With the empty space skipping technique, the network query at the location where the volume density value is zero, indicating that this space is essentially empty, could be automatically ignored for that sample location. Therefore, our proposed 3D NeRF drawing effect could further improve the overall rendering speed in VR by computing only the areas specifically requested by the user.

\section{System Implementation}\label{sec:setup}

To implement the proposed magic NeRF lens effects, we first develop a VR NeRF system. Since our technique requires combining NeRF rendering with other types of 3D representation in the conventional rasterization pipeline, we developed immersive-ngp \cite{28-Immersive-NGP,93-interactive-nerf}, the implementation of the VR-NeRF framework in the Unity game engine. 

\subsection{The VR-NeRF Plugin for Unity} \label{sec:immersive-ngp-review}

\noindent The early version of immersive-ngp that we have released includes a basic Unity native rendering plugin that supports data sharing between the instant-ngp application\cite{27-Instant-NGP} and the Unity \cite{28-Immersive-NGP, 93-interactive-nerf} render engine, allowing pre-trained NeRF models from instant-ngp to be loaded and rendered directly into the Unity game engine.  For rendering stereoscopic NeRF models within a VR HMD, two render textures are natively created and placed as screen space overlays in front of the user's eyes. Synchronization problems are avoided by updating both render textures simultaneously in a single frame.  Unity's tracked VR head poses are converted directly to instant-ngp camera matrices in the instant-ngp coordinate system. Users can navigate the 3D NeRF space using conventional VR joystick controls. In terms of software, the immersive-ngp framework uses Unity Editor version 2019.4.29f1 with the OpenVR desktop runtime, which supports a wide range of VR runtimes such as Oculus and SteamVR. Both immersive-ngp and instant-ngp use OpenGL as their graphics application programming interface (API) and support Deep Learning Super Sampling (DLSS). In this work, we use the same software architecture and extend the preliminary implementation so that a NeRF model can be transformed, manipulated, edited, and integrated as additional 3D objects into the Unity rendering pipeline. 

\subsection{Immersive-NGP Extensions} \label{sec:immersive-ngp-extension}

\paragraph{Manipulating a NeRF Model as an Object} 

\begin{figure}
    \centering
    \includegraphics[width=1\linewidth]{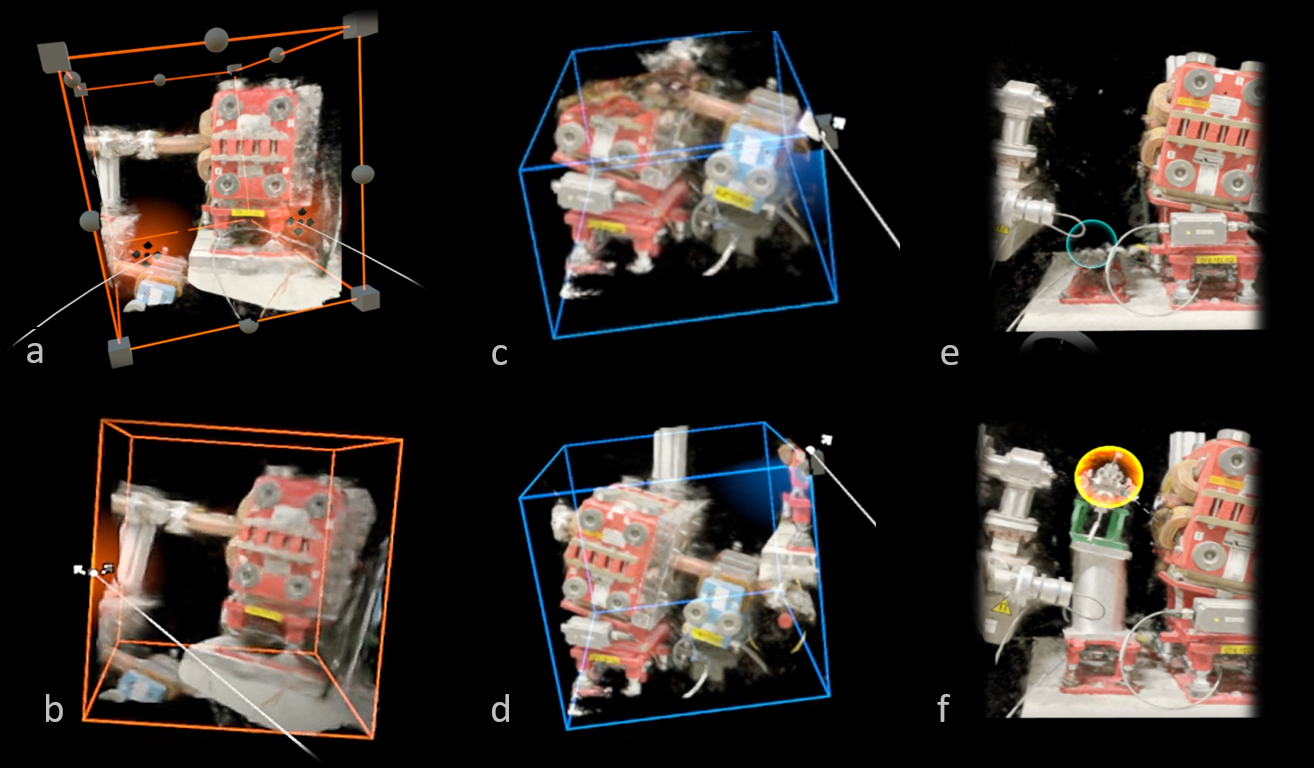}
    \caption{Illustration of our system extension to immersive-ngp. (a-b): NeRF model manipulation, where users can rotate, scale, and translate the model in VR. (c-d): NeRF model crop box manipulation, where users can rotate and scale the render volume crop box in VR. (e-f): Volume editing, where users could dynamically delete and remove an area of the render volume via 3D VR drawing.}
    \label{fig:system_extension}
\end{figure}

To support spatial manipulation of a NeRF model in VR, we create a model space for the NeRF model whose spatial properties such as position, rotation, and scale are defined by a volume bounding box. As Figure \ref{fig:system_extension} (a-b) shows, the bounding box is represented as a transparent cube whose translation ($T$), rotation ($R$), and scale ($S$) matrices are combined into one transformation matrix that is applied to the view matrix of the instant-ngp camera to render the correct view of the NeRF model using the classical view model projection theory. The object manipulation interaction from the Mixed Reality Toolkit (MRTK) is attached to the cube, allowing the user to intuitively rotate, scale, and translate the box and its associated NeRF model using one- or two-handed control. 

\paragraph{Crop Box Editing} To allow arbitrary spatial transformation of the axis-aligned bounding box (AABB) in VR, we attach a second 3D cube game object with an object manipulator where the spatial transformations of the Unity AABB bounding box, such as translation and rotation, are applied to the AABB crop box defined in instant-ngp's coordinate system.  As shown in Figure \ref{fig:system_extension} (c-d), users can extend, reduce, or rotate the AABB render volume using the same object manipulation interaction provided by MRTK. 

\paragraph{Volumetric Editing} We adopt instant-ngp's implementation of a volumetric editing feature in our framework. Similar to instant-ngp's preliminary implementation, a 3D sphere object is attached to the controller at a certain distance to indicate the intended drawing region. As Figure \ref{fig:system_extension} (e) shows, users can interactively select the region where voxels within the sphere should be made transparent by moving the sphere to the target region in 3D space while pressing a button on the VR controller to confirm the drawing action. The same interaction and action could be performed to reveal the volume at the region set to transparent, as shown in Figure \ref{fig:system_extension} (f). In an instant-ngp NeRF model, the volume density grid $\sigma$ stores the transparency of the scene learned by the network. A binary bitfield $bit_{\sigma}$ can be generated based on a transparency threshold and can be used to quickly mask the transparent region from being sampled during raytracing \cite{27-Instant-NGP}. 

\begin{figure*}
    \centering
    \includegraphics[width=1\linewidth]{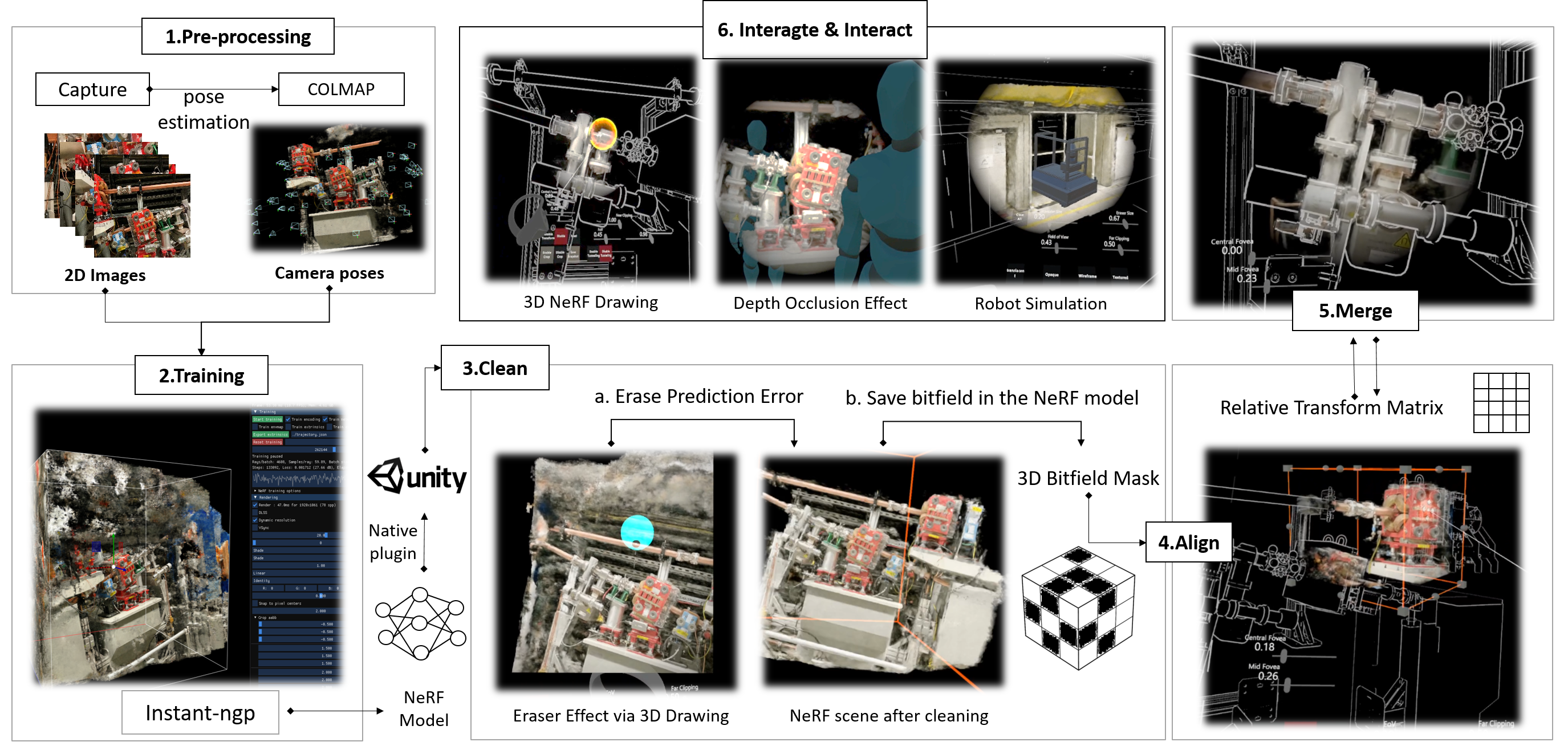}
    \caption{Overview of the data fusion pipeline to merge a NeRF model with its corresponding CAD model, illustrating sub-processes from image pre-processing, NeRF model training, scene cleaning, scene alignment, scene merging, and examples of final integration and interaction using different features of our framework.%
    }
    \label{fig:fusion_overview}
\end{figure*}

\paragraph{Depth-based Rendering} Depth-based rendering is important for creating applications that need to incorporate various visual effects such as occlusion, physics simulation, and scene relighting \cite{54-depth-lab-google} for image-based rendering. We integrate depth-based rendering into the Unity game engine by implementing an additional CUDA surface object in instant-ngp, where the depth map from the NeRF camera can be copied directly to a Unity texture on the GPU. When enabled, the per-eye depth map of the NeRF camera could be rendered at every frame, and users could select different depth scales according to the needs of their application. As shown in Figure \ref{fig:fusion_overview}, a depth occlusion effect could be achieved to integrate virtual objects into the NeRF model by performing a pixel-by-pixel depth test between the Unity camera depth map and the NeRF camera depth map. \\

In summary, the immersive-ngp extensions introduced in this work provide many core components for building interactive VR NeRF applications. The preliminary implementations of the work have already been presented at two VR demo exhibitions \cite{28-Immersive-NGP, 93-interactive-nerf}, where users could test the end-to-end pipeline of scanning, manipulation, and editing. A full demo video of the current system can be found in the supplementary material.

\subsection{Data Fusion Pipeline for Multi-modal Rendering} \label{sec:data-fusion}

\noindent In this section, we focus on developing a data fusion pipeline to merge a NeRF model with its corresponding CAD model, although the proposed pipeline is also applicable to other types of conventional 3D data. We choose CAD models because CAD models of large-scale physics facilities are often created in the early stages of facility design and have complementary features to their corresponding NeRF models. In contrast to industrial product modeling, CAD models of large-scale industrial facilities typically lack realistic textures because the detailed environments are usually too complex to be modeled accurately and the environment is bound to be updated frequently \cite{82-VR-factory-point-cloud}. In addition, semantic information typically embedded in CAD models is not represented in NeRF models without additional expensive network training and voxel-wise segmentation. Other benefits of fusing a CAD model with a NeRF include a more efficient depth occlusion effect using the CAD model without the need to perform pixel-wise depth testing, while potentially laying the foundation for more advanced NeRF editing interactions such as those proposed in NeRFShop \cite{33-NeRFshop23}.

Figure \ref{fig:fusion_overview} provides an overview of the data fusion pipeline, which is divided into the following six steps. 

\begin{enumerate}
    \item \textbf{Preprocessing.} Since most 2D images do not contain their camera poses, the 2D images must be preprocessed using the conventional Structure from Motion (SfM) algorithm \cite{56-SFM} to estimate the camera poses using software such as COLMAP \cite{55-COLMAP}. However, this step can be skipped for cameras that can track their own poses. 
    
    \item \textbf{Training.} The processed data is trained in the instant-ngp framework to generate an initial estimate of the scene function $F_{\Theta}$ and an initial occupancy grid $\sigma_0$ whose size is defined by a preset AABB bounding box.
    
    \item \textbf{Scene cleaning.} The quality of NeRF rendering can be affected by artifacts of real-world 2D images, such as motion blur, lens distortion, or insufficient images around the viewing angle. This often results in false clouds in the 3D NeRF scene, which can degrade the user's viewing experience and needs to be cleaned up before the NeRF model can be applied to practical applications. Therefore, our VR NeRF framework provides the ability to modify the pre-trained instant-ngp NeRF model $F_{\theta}$. The initial density grid estimation $\sigma_0$ is inspected and cleaned using the interactive eraser function described in section \ref{sec:immersive-ngp-extension}. Users can manually remove regions with cloudy prediction errors or low render quality. The edited density grid $\sigma_1$ and the binary bitmask of the density grid $\sigma_{bit}$ are stored in the NeRF model for reuse. When reloading the cleaned NeRF model, our framework automatically checks the saved bitmask, so that the erroneous network prediction that has already been removed will no longer be sampled and rendered. 

    \item \textbf{Scene Alignment.} The user could manually align the cleaned NeRF model with the CAD model using the NeRF object manipulation and crop box editing functions described in section \ref{sec:immersive-ngp-extension}. If the NeRF model is adjusted to the one-to-one scale of the real facility, alignment using the entire NeRF model will result in a rendering volume too large to be efficiently manipulated in VR. Therefore, it is recommended to perform the alignment process by focusing on a small part of the scene for accurate object manipulation.
    
    \item \textbf{Scene Merging}. In the merge step, the user could perceptually validate the alignment by reducing the FoV of the NeRF camera and adjusting the transparency of the shader that renders the NeRF images so that both the CAD drawing and the NeRF rendering are simultaneously visible. The user can iteratively move back and forth between the alignment and fusion steps until the two 3D representations are spatially aligned. The relative transformation matrix between the NeRF model and the CAD model can be saved for reuse.
    
\end{enumerate}

\section{Performance Benchmarking}\label{sec:performance_banchmark}



\begin{figure*}
    \centering
    \includegraphics[width=1\linewidth]{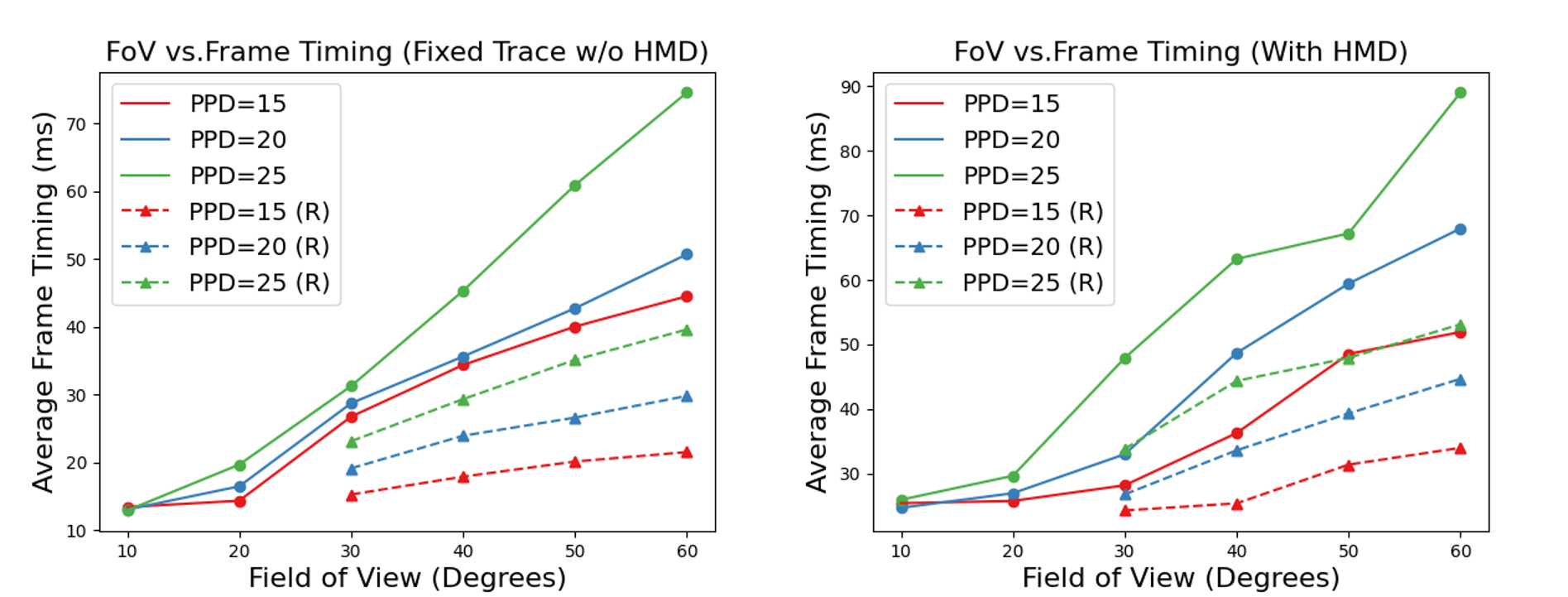}
    \caption{Systematic benchmark results showing (left): the relationship between rendering FoV and average frame time following a predefined fixed path without a VR HMD attached, and (right): the trend between rendering FoV and average frame timing for a test user following only approximately the same 3D trace. (R) indicates reduced rendering via 3D NeRF drawing.}
    \label{fig:performance1}
\end{figure*}

\subsection{Experiment Design} \label{sec:experiment_design} 

\noindent We examine the performance trend when varying the NeRF rendering FoV from $10 ^{\circ}$, which covers the foveal region of human vision, to $60 ^{\circ}$, which covers the average central visual field for most people \cite{78-human-eye}. We also vary the PPD value to match the resolution requirements of VR displays of different quality, including medium display resolution (PPD=$15$, e.g. Oculus Quest 2), medium to high-end displays (PPD=$20$, e.g. Meta Quest Pro), or high-end displays (PPD=$25$, e.g. Varjo XR-3). The final NeRF render resolution per eye ($R \times R$) for each configuration is calculated using the following equation:
\begin{equation}
    R = FoV \times PPD \times 2
\end{equation}
where we multiply the pixel density per degree by 2 to account for the upsampling required for aliasing reduction and edge smoothing via supersampling \cite{79-supersampling}. 

\subsection{Materials} \label{sec:experiment_material} 
\noindent We evaluate the performance of our system on a self-generated real-world dataset consisting of 60 images reconstructing a section of the EuXFEL particle accelerator. Its VR NeRF reconstruction is shown in Figure \ref{fig:study_conditions}. Since the CAD model is designed to be at an exact scale of the real facility, we aligned the NeRF model with the CAD model using the data fusion pipeline described in section \ref{sec:data-fusion} and were able to measure that the NeRF rendered volume of this NeRF model corresponds to approximately 2.2 m $\times$ 1.47 m $\times$ 2.08 m in real scale, which would require a full FoV VR rendering to examine closely.  All of our benchmarks are run on an Nvidia Geforce RTX 3090 GPU and an Intel(R) Core(TM) i7-11700K CPU with 32GB of RAM. As demonstrated in previous research \cite{28-Immersive-NGP}, DLSS is critical to achieving real-time immersive VR NeRF rendering performance, so DLSS is also enabled for all benchmarking efforts.

\subsection{Systematic Benchmark (w/o HMD)}\label{sec:systematic benchmark}
\noindent
To simulate the system performance of how a user would use our framework to inspect different parts of the NeRF model, we first asked a test user to closely inspect three fixed locations within the NeRF model. A custom script recorded the user's exact 3D trajectory with the camera transformation information at each frame. The relative position of the three target components with respect to the CAD model was visualized as a schematic sketch, as shown in Figure \ref{fig:study_conditions}-(Instruction Card). A video recording of the 3D trace is also available in the supplementary material.

First, we performed a systematic benchmark where we collected the average frame timing ($\overline{f_t}$) for stereoscopic rendering of different display FoV and PPD configurations by replaying the same pre-recorded 3D trace. Figure \ref{fig:performance1} (left) plots the average frame timing result for each rendering configuration. As expected, the rendering latency scales linearly with increasing FoV for the basic NeRF lens effect. Additionally, it is shown that our system could theoretically achieve real-time VR rendering of 30 frames per second ($\overline{f_t} < 33.33 ms$) at less than $50^{\circ}$ FoV. 

To evaluate whether additional contextual rendering could lead to real performance gains, we create an edited NeRF model by 3D drawing interaction, where we define the 3D density grid so that only the target components are visible. Since FoV settings below $20 ^{\circ}$ could already achieve good real-time performance without reducing the rendering volume, we only examine FoV and IPD settings above $30 ^{\circ}$ in this section. Figure \ref{fig:performance1} (left) also plots the average frame timing for each rendering configuration for rendering the selected regions. As expected, exposing only the target render volume reduces the overall render load due to instant-ngp's empty space skipping and early ray termination implementation. This confirms the performance optimization insight that removing unimportant scene details using the 3D NeRF drawing effect could be an option to gain additional performance.

\begin{figure*}
    \centering
    \includegraphics[width=1\linewidth]{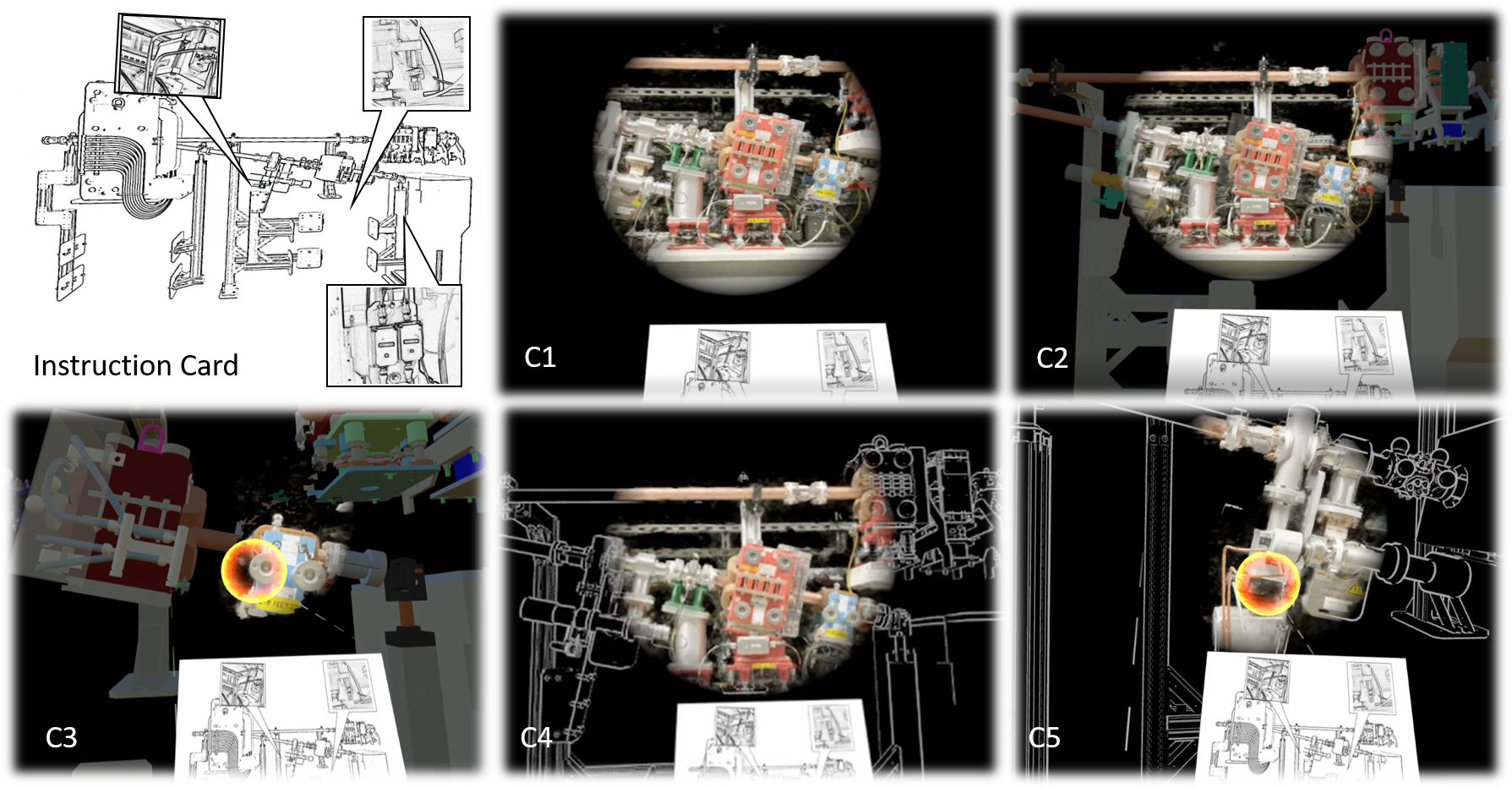}
    \caption{Screenshots for different visualization and interaction conditions of Magic NeRF Lens. (C1): Baseline condition with only the NeRF model and the FoV restrictor. (C2): Magic NeRF Lens with a textured CAD model as context. (C3): 3D NeRF drawing with a textured CAD model as context. (C4): Magic NeRF Lens using only the wireframe representation of the CAD model. (C5): 3D NeRF drawing with wireframe representation of the CAD model.}
    \label{fig:study_conditions}
\end{figure*}

\subsection{Empirical User Benchmark Results (With HMD)}\label{sec:systematic benchmark}
\noindent 
In the second benchmarking experiment, we evaluate the same set of rendering configurations. However, instead of replaying a pre-recorded 3D trace, a test user was asked to repeat each configuration, following roughly the same path to inspect the three target components to evaluate the overall performance of the actual VR system. For high-resolution configurations where real-time performance could not be achieved, the test user was able to use keyboard control instead, while keeping the VR HMD connected to the entire rendering pipeline. Figure \ref{fig:performance1} (right) shows the general trend between FoV and $\overline{f_t}$. Compared to the systematic performance results, the actual total frame timing is about $5-10$ ms higher because of the additional computational resources needed to drive the HMD. However, the general performance trend in an actual system with VR HMD is similar to the systematic benchmark result. Additionally, we could derive that for an actual VR NeRF system, the optimal render resolution is $1200 \times 1200$ at $30^{\circ}$ FoV (PPD=20) or at $40^{\circ}$ FoV (PPD=15).

\section{User Study}

\noindent Understanding human factors is an important step in the development of VR systems and interaction techniques. Moreover, the impact of using NeRF in an interactive VR framework on actual users is currently unknown and unexplored in the community. In this section, we present a perceptual experiment where we invited 15 users to perform a facility inspection task through visual search activities using different magic NeRF lens effects. We collect various perceptual metrics to report on users' perceived motion sickness, sense of presence, task load, and perceived usability, from which we derive design insights for future applications. 

\subsection{Study Design} \label{sec:conditions}

\paragraph{Conditions} In this user study, we are interested in the effects of different magic NeRF lenses on the performance of virtual facility inspection tasks. In particular, we focus our study design on the perceptual effects of fusing NeRF with its corresponding CAD models. We investigate the two magic lens designs with the two most common types of CAD model visualizations in large-scale industrial facilities. As shown in Figure \ref{fig:study_conditions}, (C1) is the baseline condition with only FoV restriction but no data fusion for comparison, (C2, C4) implement the magic NeRF lens effect with FoV restriction, and (C3, C5) implement the magic NeRF lens effect with 3D drawing. (C2, C3) use a CAD model with colored abstract texture, while (C4, C5) use a CAD model with only wireframe representation, visualizing only the contextual outline. 

\paragraph{Materials} The user study was conducted on the same graphics workstation and software configuration used for the system benchmark described in section \ref{sec:performance_banchmark}. For all conditions, the FoV of the NeRF rendering camera was set to $30 ^{\circ}$ with a per-eye render resolution of $1200 \times 1200 $ pixel, which is one of the optimal rendering configurations determined in our benchmark experiments. For the 3D NeRF drawing conditions (C3) and (C5), the reveal sphere is placed at $0.7 m$ in front of the right VR HMD controller. We used an Oculus Quest Pro VR HMD, which has a PDD value of 22. However, our VR NeRF framework is compatible with other VR headsets that support the SteamVR and OpenVR desktop runtime.  

\paragraph{Tasks} To simulate how a user would perform virtual inspection tasks, we designed a visual search activity where the user was asked to locate three detailed components in the NeRF model based on a schematic sketch. As shown in Figure \ref{fig:study_conditions}-(Instruction Card), the schematic sketch consisted of an overview of the CAD model, with each search target highlighted in a dialog box. Arrows are provided to indicate an approximate location where the detailed component might be found. The instructions for each search target were generated from the real-world images and therefore represent updated real-world conditions compared to the abstract CAD model. All search targets are detailed elements that could not be effectively updated in the CAD model due to maintenance activities or the difference between actual installation and original design. A pilot test with 3 users confirmed that all target components in each scene could be found in 3-5 minutes for all proposed conditions. To avoid learning effects in the course of the study, 5 different scenes were prepared, each showing a different part of the actual particle accelerator tunnel at the EuXFEL. 




\paragraph{Participants} We invited 15 participants, 4 female, 10 male, and 1 who preferred not to disclose their gender. 5 participants were between 18 and 24 years old, and 10 participants were between 25 and 34 years old. All were students or researchers in HCI, Computer Science, or Physics, with HCI students receiving compensation in the form of course credits. 8 participants use VR systems regularly (at least once a month), and only 2 use them less than once a year, 5 never used VR before.

\paragraph{Procedures}


\noindent Each participant first signed a consent form and completed a demographic questionnaire. The experimenter then presented a test VE in which participants could practice locomotion and 3D VR drawing controls for a maximum of 3 minutes. Each participant was then presented with the 5 conditions in order. To counterbalance the order and assignment of scenes to conditions, we used a replicated Latin square design with one treatment factor (condition) and three blocking factors (participant, trial number, and scene). As a result, each condition/scene combination was experienced by 3 participants. 
During each condition, participants could press a "start" button to begin the study and begin time recording to assess task performance. The experimenter monitored the participants' VR interactions via a secondary screen that mirrored the VR displays. After the participant informed the experimenter when they had found a component, the experimenter confirmed its correctness, after which the participant could begin searching for the next search target. Once all three components were found, the participant could press an "end" button, which marked the end of time recording for the search tasks. The total study time was approximately 60 minutes.

\subsection{Results}

\begin{figure*}[t!]
    \centering
    \includegraphics[width=1\linewidth]{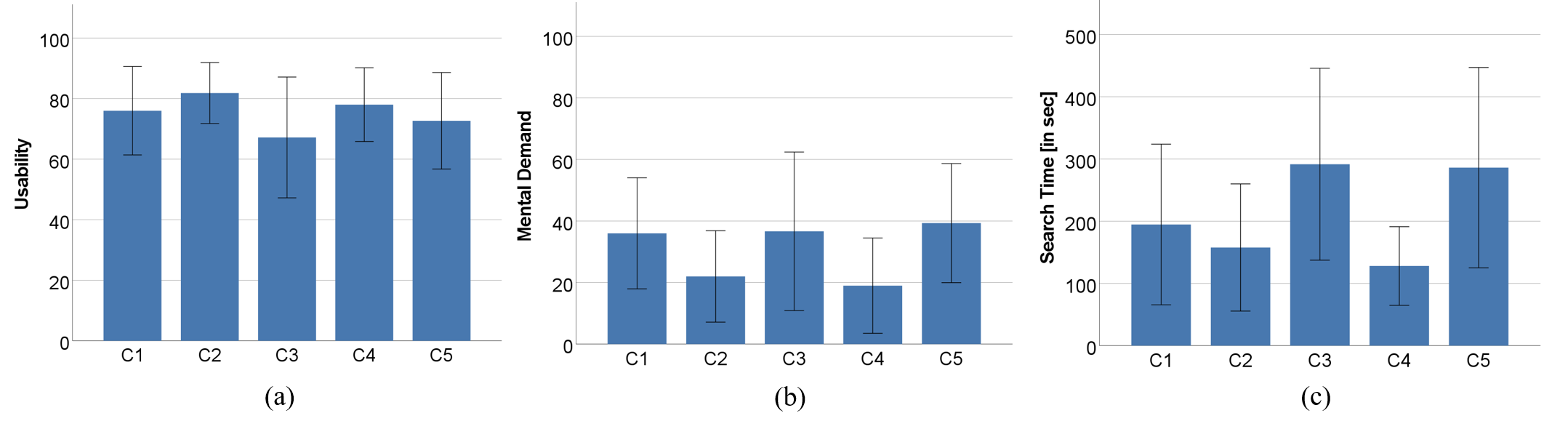}
    \caption{Mean (a) usability, (b) mental demand as measured by the NASA-TLX, and (c) search time to finish the task in seconds. Vertical bars indicates the standard deviation.}
    \label{fig:user_study_result_diagram}
\end{figure*}

\noindent We collected multiple objective and subjective measures to assess the user experience as study participants interacted with our magic NeRF lens framework. Tables listing the overall ratings for each subscale of the rating scales are included in the supplementary material. 

\paragraph{Task Performance}
Task performance was measured as the time it took participants to find the three locations indicated on the instruction card for each scene.
A Shapiro-Wilk test indicated that the residuals of search time were not normally distributed, which was confirmed by visual inspection of the QQ plots.
Therefore, a Friedman test was performed, which revealed a significant effect of visualization method on search time ($\chi^2(4)=23.307, p<.001$).
Post-hoc Wilcoxon signed-rank tests with Bonferroni-Holm adjustment of p-values showed significant differences between C2 and C3 ($Z=-3.010, p=.018$), C2 and C5 ($Z=-3.124, p=.014$), C4 and C3 ($Z=-3.237, p=.011$), and C4 and C5 ($Z=-3.351, p=.008$).

\paragraph{Perceived Workload}
We measured six aspects of perceived workload using the NASA-TLX questionnaire~\cite{73-NASA-TLX}. Because each aspect was represented by a single Likert scale ranging from 0 to 100 (with 21 levels), we used nonparametric Friedman tests to analyze the responses. There was a statistically significant difference in mental demand depending on the condition, $\chi^2(4)=17.857, p=.001$.
Post-hoc analysis with Wilcoxon signed-rank tests and Bonferroni-Holm p-value adjustment revealed a significant difference between C2 and C5 ($Z=-3.257, p=.011$).
Friedman tests also showed significant effects of visualization method on effort ($\chi^2(4)=15.366, p=.004$) and frustration ($\chi^2(4)=10.183, p=.037$), although none of the post-hoc tests were significant after p-value adjustment.
No significant differences were found for physical demand, temporal demand, and performance.

\paragraph{Usability}
Usability was measured using the System Usability Scale (SUS)~\cite{70-SUS} and converted to a value range [0,100].
A repeated measures ANOVA with Greenhouse-Geisser correction revealed a significant effect of visualization method on usability,\linebreak $F(2.514,35.202)=4.269, p=.015, \eta_p^2=.234$.
Post-hoc tests with Bonferroni correction showed a significant difference between C3 and C4 ($p=0.48$).

\paragraph{Presence}
We measured the sense of presence using the Igroup Presence Questionnaire (IPQ)~\cite{71-IPQ} with all three subscales \textit{Spatial Presence}, \textit{Involvement}, and \textit{Experienced Realism}, as well as a single item assessing the overall "sense of being there". Three ANOVAs for the subscales and a Friedman test for the single item revealed no significant differences.

\paragraph{Cybersickness}
As a subjective measure of cybersickness, participants rated the perceived severity of 16 symptoms covered by the Simulator Sickness Questionnaire (SSQ)~\cite{72-SSQ}. From the ratings, we calculated subscores for nausea, oculomotor, and disorientation, as well as a total cybersickness score, as suggested by Kennedy et al.~\cite{72-SSQ}.
No increase in cybersickness across trials was observed, so we analyzed absolute values rather than relative differences between trials.
Since the residuals were not normally distributed, we performed a Friedman test for each of the four (sub)scores.
For nausea, a significant difference between conditions was found, $\chi^2(4)=10.330, p=.035$, but this could not be confirmed by post-hoc pairwise Wilcoxon signed-rank tests with Bonferroni-Holm correction.
For oculomotor, disorientation, and total cybersickness scores, no significant effects were found using Friedman tests.


\subsection{Discussion}

\paragraph{Spatial Orientation} 
The magic lens effects with FoV restriction (C2, C4) achieved better ratings in terms of mental effort and higher task performance than C1, though not being significant. Qualitative user feedback indicated that C2 and C4 ``\textit{make(s) it more confident /easier to navigate around the machines}`` (N=2). Without the contextual guidance of CAD models, ``\textit{it was difficult to get a feeling of scale or to identify different places correctly.}" Detailed comments mentioning spatial orientation confirmed that these effects further ``\textit{helped with general orientation}`` and ``\textit{provided a sense of [the user's] position}``. 
In contrast, the 3D NeRF drawing interaction (C3, C5) performed significantly worse than the 3D NeRF drawing effect (C2, C4). The main problem reported with these conditions was ``\textit{difficulty to find the correct depth}`` (N=4). This could be because most participants were unfamiliar with the complex environments, and using only a 2D schematic sketch as visual instruction does not provide users with sufficient spatial orientation to navigate efficiently to the search target. As a result, future design of the magic NeRF lens effect with 3D drawing could provide 3D spatial anchors as visual cues for non-expert users. 


\paragraph{Task Load, Performance, Usability, and Cybersickness}
Across all conditions, there were no significant differences in perceived cybersickness, with participants experiencing little to no motion sickness. In general, users found the magic lens effects with FoV restrictor (C2, C4) highly usable ($SUS_{C2}=81.83$, and $SUS_{C4}=78$), with better objective task performance and lower perceived mental effort. However, the magic NeRF lens effects with 3D drawing received lower usability scores, with a significant difference between (C3) and (C4). As Figure \ref{fig:user_study_result_diagram} shows, usability scores were strongly correlated with cognitive load and task performance. Qualitative feedback indicates that participants often repeatedly unfolded and erased the NeRF models because they were unsure where to find the components, leading to frustration and even framerate jitters when more areas than necessary were revealed. In addition, participants mentioned that the CAD is often outdated and incomplete compared to the NeRF model. This discrepancy makes it even more difficult for users without a facility management background to complete the tasks. Nevertheless, two participants mentioned the 3D NeRF drawing effect as their preferred condition because ``\textit{it was more fun} `` and has ``\textit{a nice property of reducing complexity and putting the focus on the spots you want to investigate}``. Therefore, we recommend future use of 3D NeRF drawing effect mainly for facility management experts, but still make it an option for non-expert users to accommodate different preferences. 

\paragraph{Context Rendering Style}
With respect to the rendering style of the CAD model (colored solid vs. wireframe), the conditions yielded similar results in both subjective ratings and task performance.
Since users reported different preferences in the open-ended feedback, a customization option could be offered in a practical application.

\section{Expert Feedback} \label{sec:experts}

\noindent To explore how our framework could be used by practitioners in a real industrial setting, we validated our system through expert reviews at DESY, where the development of a VR NeRF system for particle accelerator maintenance was first proposed. Five facility managers and control system specialists participated in the reviews. All participants have a leading position in the design, coordination or control of particle accelerators at DESY, and two of them are also experienced VR experts who have already developed VR systems for facility inspection.

The expert reviews were conducted using an exploratory application that illustrates a section of the NeRF model of the EuXFEL particle accelerator. It provides many flexibility and customization options, allowing the user to freely adjust the NeRF camera's field of view, change the size of the NeRF editing sphere, vary the translucency of the merging effects, as well as the manipulation interactions mentioned in section \ref{sec:setup}. The application ran on an Alienware m17 R2 laptop with 16GB of RAM and an RTX 2080 GPU, where we reduced the resolution of the application to 800 $\times$ 800 pixels to achieve real-time performance. 

Overall, participants felt ``\textit{very confident to use the system}'' and ``\textit{it is something (they) could work with}'', even though the application was running on a laptop with moderate performance and moderate resolution.  All expert participants confirmed that using NeRF for virtual facility inspection could benefit their workflow. One facility management expert commented that NeRF is a compelling, low-cost alternative for 3D facility documentation: ``\textit{I think the system has a good advantage. It is quite nice to project the NeRF model on the CAD model, as it is a lot more effort to take laser scans of the facility}''.  In addition, most expert participants preferred to inspect the facilities in VR because the system ``\textit{helps them to see if (they) could reach anything}'' or ``\textit{if (they) could fit any equipment through the existing environment}''. In addition, two experts mentioned that having a 1:1 scaled NeRF model aligned with the CAD model in immersive VR also gives them a better spatial awareness of the complex machines than working with a 2D desktop application. For example, they mentioned that the VR NeRF environment could help them assess in advance if ``\textit{an operator's hand would fit through a narrow gap to handle components}'' or if special equipment would need to be prepared in advance. 

The expert who leads the design and upgrade of particle accelerators mentioned the benefits of data fusion visualization: ``\textit{With this system, I see the possibility to test something in theory before you build it in practice. For example, when you have a machine, and you want to test if you have enough space for installing it, it is quite nice you could test everything in the virtual area before you do it in reality}''. He also mentioned that the CAD model of the particle accelerator usually only shows the initial design of the facility. Once the facility is operational, these CAD models can become incomplete and outdated. He found the contextual 3D NeRF drawing effect particularly helpful in comparing the difference between the original design and the actual implementation, which could even help operators update the original CAD designs accordingly.

For future development of the facility inspection system, one expert suggested the interesting idea of integrating contextual QR code scanning to further support information retrieval from their large inventory database. Sometimes, facility inspection tasks at the particle accelerator require scanning labels containing QR codes with manufacturing and maintenance information that are attached to all cables and equipments. The ability to retrieve such labels directly from the NeRF model would further streamline facility inspection processes. With regard to safety-critical processes such as immersive robot teleoperation, participants mentioned that although NeRF could provide photometrically accurate results, it's geometric accuracy also needs to be verified and compared with conventional 3D sensors. Nevertheless, as our proposed NeRF magic lens effects and data fusion pipeline could be applied to other types of conventional 3D models, we encourage further investigation of experimenting with data fusion with other types of 3D data of large-scale facilities.

\section{Limitations}

Our framework still has several limitations that could be addressed in future work. First, future research could investigate an automatic CAD-NeRF alignment approach to skip the manual hand-eye calibration process. Detailed investigation of an interactive point-matching algorithm could be a promising approach, taking into account the real-world mismatch between the NeRF model and the CAD model \cite{74-iterative-point-matching}. In addition, optimization techniques such as empty space skipping and early ray termination could also lead to frame rate jitter as the number of network queries becomes view-dependent. Although the use of a FoV restrictor could reduce such effects, such framerate jitter will be more noticeable on medium and low end graphics hardware. Therefore, we encourage future work to investigate the integration of further optimization techniques, such as foveated rendering combined with user visual saliency, to enable a comfortable VR NeRF system even on low-end graphics devices. 

\section{Conclusion}

We presented Magic NeRF Lens, an interactive immersive VR framework that can support virtual facility inspection using both photorealistic NeRF rendering and geometric rasterization. From the system performance benchmark, we show that the optimal VR NeRF rendering configuration for visualizing large-scale facilities at one to one real-world size is $20$ PPD at $30^{\circ}$ FoV, or $15$ PPD at $40^{\circ}$ FoV. To support rendering of large NeRF models, we used multimodal data fusion designs for contextual and on-demand NeRF rendering. Through a visual search user study, we demonstrate that our MR tunneling magic NeRF lens design achieves high usability and performance, while the 3D NeRF drawing effect is more interactive but more suitable for expert users. Follow-up system reviews with 5 experts confirmed the results of the user study and suggested various directions for future work. We believe that the interdisciplinary nature of this work could benefit both industrial practitioners and the VR community at large.  

\section*{Acknowledgments}

This work was supported by DASHH (HELMHOLTZ Graduate School for the Structure of Matter) with the Grant-No. HIDSS-0002, the German Federal Ministry of Education and Research (BMBF), and the German Research Foundation (DFG). 

\bibliographystyle{IEEEtran}
\bibliography{cite}
\end{document}